\documentclass[12pt,a4paper]{article}
\usepackage{fullpage}
\usepackage{graphicx,subfigure}
\usepackage{epsfig}
\newcommand{\text}[1]{\mbox{\rm #1}}
\newcommand{\eqref}[1]{(\ref{#1})}
\begin{document}
\begin{center}
{\bf Exceptional orthogonal polynomials, QHJ formalism and SWKB quantization condition}\\
S. Sree Ranjani$^a\,$\footnote{s.sreeranjani@gmail.com}, P. K. Panigrahi$^b\,$\footnote{prasanta@iiserkol.ac.in}, A. Khare$^c\,$\footnote{khare@iiserpune.ac.in}, A. K. Kapoor$^a\,$\footnote{akksp@uohyd.ernet.in} and  A. Gangopadhyaya$^d\,$\footnote{A. Gangopadhyaya}\\
$^a$School of Physics, University of Hyderabad, Hyderabad, 500 046, India.\\
$^b$Indian Institute of Science Education and Research (IISER) Kolkata, Salt Lake, Kolkata,700 106, India.\\
$^c$Indian Institute of Science Education and Research (IISER) Pune, 411021, India.\\
$^d$Loyola University Chicago, Department of Physics, Chicago, IL 60660.\\
\end{center}

\begin{center}
Abstract
\end{center}
We study the quantum Hamilton-Jacobi (QHJ) equation of the recently obtained
exactly solvable models, related to the newly discovered exceptional
polynomials and show that the QHJ formalism reproduces the exact eigenvalues
and the eigenfunctions. The fact that the eigenfunctions have zeros and poles
in  complex locations leads to an unconventional singularity structure of the
quantum momentum function $p(x)$, the logarithmic derivative of the wave
function, which forms the crux of the QHJ approach to quantization.  A
comparison of the singularity structure for these systems with the known
exactly solvable and quasi-exactly solvable models reveals interesting
differences. We find that the singularity structure of the momentum function for these new potentials lies between the above two distinct models, sharing similarities with both of them. This prompted us to examine the exactness of the supersymmetric WKB (SWKB) quantization condition. The interesting singularity structure of $p(x)$ and of the superpotential for these models has important consequences for the  SWKB rule and in our proof of its exactness for these quantal systems.


\noindent
 \section{Introduction}

New infinite sets of solvable quantum mechanical  potentials have been recently constructed by Odake and Sasaki \cite{odake},\cite{odake2}, using the Darboux-Crum transformations \cite{dar},\cite{crum}. They deformed the  standard radial oscillator and Darboux-P\"oschl-Teller potentials using the Laguerre and the Jacobi polynomial eigenfunctions of degree $\ell$.  The solutions of the new potentials are in terms of the newly discovered Laguerre- and the Jacobi-$X_{\ell}$ type  exceptional orthogonal polynomials  respectively, which have been recently introduced  by G\'omez-Ullate {\it et.al.}, \cite{nick1},\cite{nick2}. The characteristic feature of this new class of polynomials is that the  series  starts with a polynomial of degree $\ell$ $(\ell=1,2 \dots )$ and still can form a complete set with respect to a positive-definite measure. This is unlike the classical orthogonal polynomial series, which require a constant to form a complete set. In addition to this feature, the eigenfunctions of these new potentials have  singularities at complex locations, associated with the zeros of the eigenpolynomials of the standard potentials from which they are derived.  These standard polynomials seem to play a crucial role in the construction of these new potentials.  The fact that the new Laguerre- and the Jacobi-$X_{\ell}$ type polynomials occur as solutions to the Sturm-Liouville problem with rational coefficients, associated  to the standard Laguerre and Jacobi polynomials, is very intriguing. In this light the quantum Hamilton-Jacobi (QHJ) analysis \cite{lea1}, \cite{lea2}, which uses the singularity structure of the logarithmic derivative of the wave function to obtain the eigenvalues and eigenfunctions for a given potential, can reveal interesting features pertaining to these models.

In this paper, we analyze the set of potentials obtained by deforming the radial
oscillator using the quantum Hamilton-Jacobi (QHJ) formalism and investigate the
singularity structure of the logarithmic derivative of the wave function,
$\psi(x)$:
 \begin{equation}
      p(x)= -i\hbar \frac{\partial_x\psi(x)}{\psi(x)}, \label{qmf}
      \end{equation}
known as the quantum momentum function (QMF). The knowledge of the singularity
structure of the QMF, coupled with the exact quantization condition,
\begin{equation}
\frac{1}{2\pi}\oint_C p(x) dx = n\hbar,  \label{exq}
\end{equation}
defined within this formalism, allows us to arrive at
the required solutions \cite{bhallaAJP}-\cite{the}. Here, $n$ gives the number of nodes
of $\psi(x)$ and the contour $C$ encloses these nodes located in the classical
region, in between the two turning points $x_1$ and $x_2$ in the complex
$x$-plane.  In the present study, in addition to obtaining the energy
eigenvalues and eigenfunctions, we concentrate on bringing out the new features
of the singularity structure of the QMF of these potentials. This, when compared
to that of the models studied earlier, namely the exactly solvable (ES) models, whose solutions are
in terms of the classical polynomials, and the quasi-exactly solvable (QES)
models revealed interesting differences. This led us to a careful analysis of the
exactness or otherwise of the supersymmetric WKB (SWKB) quantization condition \cite{khareAJP}, \cite{seetha} for these new
models, because it was shown in \cite{bhalla} that the singularity structure of
the QMF provided a link to the exactness of the SWKB integral. Here, Bhalla {et.al.,} showed that the SWKB condition is exact for
all models, where the locations and the residues of the poles in the non-classical regions  of the SWKB integrand and the QMF matched identically. Hence, it will be interesting to see if a similar analysis works for these new  models,
especially when their singularity structure differs from that of the conventional models.

In the following section, we give a brief account of the QHJ formalism \cite{lea1}-\cite{the}, followed
by the analysis of the generalized deformed radial oscillator potential within
this method. We then compare the singularity structures of
various models, in order to identify the similarities or differences, between the ES and QES models and the present ones. In section 3 we investigate the exactness of the SWKB condition
for the deformed radial oscillators and subsequently conclude in the last section, with directions for further research.

 \noindent
\section{QHJ formalism and the deformed radial oscillators}
      In the QHJ formalism, $x$ is treated as a complex variable and we use the
Riccati equation for $q \,(=ip(x))$ $(\hbar =2m =1)$
\begin{equation}
   q^2 +\partial_x q+(E-V(x))=0,  \label{qhj}
\end{equation}
to obtain the required solutions. Using \eqref{qmf} and substituting $q=\frac{\psi^{\prime}(x)}{\psi(x)}$ in the above equation gives us the Schr\"odinger equation
 \begin{equation}
 -\frac{d^2\psi(x)}{dx^2} +V(x)\psi(x)= E\psi(x),
 \end{equation}
which establishes the connection between QHJ approach and the conventional ones. In this approach, the singularities of the momentum function $q$ play the key role in obtaining the spectra, without the need to solve for the eigenfunctions.

   The singularities of $q$ are of two types namely, the fixed and the moving
   singularities. The fixed singularities correspond to the singularities of
   the potential and their location is independent of the energy and can be
   located by inspection from (\ref{qhj}). The moving singularities are poles, which
   correspond to the  zeros of the wave function $\psi(x)$. It is well known
   that the location of the nodes of the wave function changes with
   energy. Therefore, the location of these poles depends on the initial
   condition and hence they are named moving poles. The number and the location of these poles cannot be
inferred from the differential equation by inspection. These poles turn out to be finite in number for all the ES models studied \cite{es}, \cite{the}. This is equivalent to the fact that the point at
infinity, at best, is an isolated singularity. The location of all these singularities and
their residues are used to evaluate the integral in \eqref{exq}, which gives the
expression for the eigenvalues for the ES models \cite{es} and the quasi-exact
solvability condition for the QES models \cite{qes}. In addition, the knowledge
of the singularity structure of $q$ allows us to obtain the form of QMF and hence the
eigenfunctions using \eqref{qmf}. For more details,  we refer the reader to our earlier papers
\cite{bhallaAJP}-\cite{the} and we proceed to analyze the
potentials under study.\\

\noindent
{\bf Deformed radial oscillator potentials}\\
Here, we present a brief account of the set containing infinite number of  new
shape invariant potentials (SIPs) $V_{\ell}(x)$, indexed by  $\ell$ $(\ell=1,2,....)$ described in \cite{odake}, \cite{odake2}. The potentials $V_{\ell}(x)$ constitute the hierarchy of supersymmetric
potentials and are translationally shape invariant \cite{khareAJP}, \cite{khbook}. Moreover, within
each set, putting $\ell=0$ gives the respective standard potential and $\ell=1$
corresponds to the potentials constructed by Quesne \cite{quesne}. In \cite{asim}, all the SIPs including the newly deformed potentials have been obtained from  a general analysis of shape invariance requirement.  Here, we analyze the set of potentials obtained by deforming the standard radial oscillator. For every potential in this set, indexed
by $\ell$, a prepotential
 \begin{equation}
\omega_{\ell}(x;g)=\omega_0(x;g+\ell)+\ln
\frac{\xi_{\ell}\,(x^2;g+1)}{\xi_{\ell}\,(x^2;g)},\,\,\,\, g > 0 \label{prepot}
\end{equation}
is defined, where
\begin{eqnarray}
\omega_0(x;g)&=&-\frac{1}{2}x^2+g\log x, \, \,\, 0<x<\infty ,\\
\xi_{\ell}(x^2;g)&=&L_{\ell}^{(g+\ell-\frac{3}{2})} (-x^2).  \label{xil}
\end{eqnarray}
Here $L_{\ell}^{(g+\ell-\frac{3}{2})} (x)$ is the associated Laguerre
polynomial. The corresponding superpotential is given by
\begin{equation}
W_{\ell}\,(x) = -\partial_xw_{\ell}(x;g) =
x-\frac{g+\ell}{x}-\frac{\partial_x\xi_{\ell}(x^2;g+1)} {\xi_{\ell}(x^2;g+1)}+\frac{\partial_x\xi_{\ell}(x^2;g)} {\xi_{\ell}(x^2;g)} \label{Gsuppot}
\end{equation}
and the corresponding potential, $V_{\ell}(x)$, with zero ground state energy is given by
\begin{equation}
V_{\ell}(x)= (\partial_x w_{\ell}(x;g))^2+\partial_{xx} w_{\ell}(x;g).
\end{equation}
Substituting $\omega_{\ell}(x;g)$  in the above equation and simplifying we obtain,
\begin{eqnarray}
V_{\ell}(x) &=& U^2_0(x) +\partial_xU_0(x)+ \frac{\partial_{xx}\xi_{\ell}(x^2;g+1)} {\xi_{\ell}(x^2;g+1)} \nonumber \\ &&+2\left(-x+\frac{g+\ell}{x}- \frac{\partial_x\xi_{\ell}(x^2;g)} {\xi_{\ell}(x^2;g)} \right)\frac{\partial_x\xi_{\ell}(x^2;g+1)} {\xi_{\ell}(x^2;g+1)},   \label{vint}
\end{eqnarray}
where
\begin{equation}
U_0(x) = -x+\frac{g+\ell}{x} - \frac{\partial_x\xi_{\ell}(x^2;g)}{\xi_{\ell}(x^2;g)}.
\end{equation}
From the differential equation and the recurrence relations of the associated Laguerre polynomials \cite{morse},\cite{odake2}, we obtain
\begin{equation}
\frac{\partial_{xx}\xi_{\ell}(x^2;g+1)} {\xi_{\ell}(x^2;g+1)}=-2\left(-x+\frac{g+\ell}{x}\right)\frac{\partial_x\xi_{\ell}(x^2;g+1)} {\xi_{\ell}(x^2;g+1)}+4\ell    \label{appen1}
\end{equation}
and
\begin{equation}
 \left(\frac{\partial_x\xi_{\ell}(x^2;g)}{\xi_{\ell}(x^2;g)}\right)\left( \frac{\partial_x\xi_{\ell}(x^2;g+1)} {\xi_{\ell}(x^2;g+1)}\right)=2x\frac{\partial_x\xi_{\ell}(x^2;g+1)} {\xi_{\ell}(x^2;g)}\left(1-\frac{\partial_x\xi_{\ell}(x^2;g)} {\xi_{\ell}(x^2;g+1)}\right)
\end{equation}
respectively. The use of these equations in \eqref{vint}, reduces the potential to
\begin{eqnarray}
V_{\ell}(x) =  U^2_0(x) +\partial_xU_0(x) -4x\frac{\partial_x\xi_{\ell}(x^2;g+1)} {\xi_{\ell}(x^2;g)} +4\ell.
\label{Gvminus}
\end{eqnarray}
Substituting \eqref{Gvminus} in \eqref{qhj} gives us the  QHJ equation
\begin{eqnarray}
q_{\ell}^2 + \partial_xq_{\ell}+ E=  U^2_0(x) +\partial_xU_0(x) -4x\frac{\partial_x\xi_{\ell}(x^2;g+1)} {\xi_{\ell}(x^2;g)} +4\ell.   \label{Gqhj}
\end{eqnarray}
Here, $q_{\ell}$ corresponds to the momentum function associated to $V_{\ell}(x)$. From the above equation we can see that  $q_{\ell}$ has a simple
pole at the origin along with $2\ell$ fixed poles corresponding to the zeros of
$\xi_{\ell}(x^2;g)$.  An important feature of all known ES models studied has been that the point at infinity is an isolated singularity \cite{es}. The consequence of this is that the QMF has finite number of moving poles. For the new ES models under study we make the same assumption, {\it i.e.,} momentum function has a finite number, $N$, moving poles.  From \eqref{Gqhj}, we can see that $q_{\ell}$ behaves like $x$ at infinity. This coupled with the knowledge of its singularities and residues allows us to write $q_{\ell}$ in the form of a meromorphic function
\begin{equation}
q_{\ell}=-x + \frac{g+\ell}{x}-\sum_{i=1}^{2\ell} \frac{1}{x-a_i}+\sum_{j=1}^N \frac{1}{x-b_j}
+ \phi(x).  \label{qmfmero}
\end{equation}
The first term on the right hand side of the above equation gives the large $x$
behaviour of $q_{\ell}$ and the second term is the
singular part arising from the  pole at the origin. The numerator is the residue
at the origin and is calculated by substituting the Laurent expansion of $q_{\ell}$
around the same :
\begin{equation}
q_{\ell}=\frac{b_1}{x} +a_0+a_1 x+......  \label{lau}
\end{equation}
in \eqref{Gqhj}. Equating the coefficients of the same powers of $x$ to zero, we
obtain two values for $b_1$, namely $\pm (g+\ell)$. The correct value is chosen using the boundary
condition, in the limit $E \rightarrow 0$, $ q_{\ell} \rightarrow -W_{\ell}(x)$. This gives  the residue to be $(g+\ell)$.
The third term in
\eqref{qmfmero} gives the singular part associated with  $2\ell$ fixed poles, with residue $-1$,
corresponding to the zeros of the $\xi_{\ell}(x^2,g)$. It is easily seen that this is equivalent to   $\frac{\partial_x\xi_{\ell}(x^2,g)}{\xi_{\ell}(x^2,g)}$.  The fourth term corresponds to $N$ simple moving poles with residue $1$. Some of these are located on the real line and others off the real line. The ones which lie on the positive real line correspond to the nodes of the wave function. The contribution of the moving poles of $q_{\ell}$ is similarly written as $\frac{\partial_xP_N(x)}{P_N(x)}$, with $P_{N}(x)$ being a polynomial of degree $N$. The last term, $\phi(x)$, describes the analytic part of $q_{\ell}$. From Liouville's theorem, $\phi(x)$ is a constant, which turns out to be zero from \eqref{Gqhj}. Substituting \eqref{qmfmero} in \eqref{Gqhj} and after a few algebraic manipulations
we get
\begin{eqnarray}
\frac{\partial_{xx}P_N(x)}{P_N(x)} + 2\left(-x+\frac{g+\ell}{x}- \frac{\partial_x\xi_{\ell}(x^2;g)} {\xi_{\ell}(x^2;g)}
\right)\frac{\partial_x P_N(x)}{P_N(x)}\nonumber\\+4x\frac{\partial_x\xi_{\ell}(x^2;g+1)} {\xi_{\ell}(x^2;g)}-4\ell +E=0. \label{temp}
\end{eqnarray}
For large $x$ the leading term in the above equation is a constant, therefore equating
it to zero, we obtain
\begin{equation}
E= 2N-4\ell.  \label{energy1}
\end{equation}
Substituting this in \eqref{temp}, we obtain
 \begin{eqnarray}
\partial_{xx}P_N\,(x) + 2\left[-x+ \frac{g+\ell}{x} -
\frac{\partial_x\xi_{\ell}(x^2;g)} {\xi_{\ell}(x^2;g)}\right]\partial_xP_N(x) + \nonumber\\
\left(4x \frac{\partial_x\xi_{\ell}(x^2;g+1)} {\xi_{\ell}(x^2;g)}+ 2N-8\ell \right)P_{N}\,(x) = 0. \label{fdiff}
\end{eqnarray}
The above equation can be written as
\begin{equation}
\frac{1}{{\bf w^2_\ell}}\frac{d}{dx}\left({\bf w^2_{\ell}}(x)\frac{d}{dx}P_N(x)\right) \\
+ \left( -4x\frac{\partial_x\xi_{\ell}(x^2;g+1)} {\xi_{\ell}(x^2;g)}+2N-8\ell\right)P_N(x) = 0   \label{tempwt}
\end{equation}
where,
\begin{eqnarray}
{\bf w_{\ell}}(x)&=& \exp\left[\int \left(-x +\frac{g+\ell}{x} - \frac{\partial_x\xi_{\ell}(x^2;g)} {\xi_{\ell}(x^2;g)} \right) dx \right],\nonumber \\
&=& x^{(g+\ell)}\exp(-\frac{1}{2}x^2)/\xi_{\ell}(x^2;g). \label{wtf}
\end{eqnarray}
The differential equation \eqref{tempwt} implies that ${\bf w_{\ell}^2}$ is the weight function with respect to which the polynomials, $P_N(x)$, are orthogonal. The function ${\bf w_{\ell}^2\,(x)}$, coincides with the weight function, associated with the exceptional Laguerre polynomials \cite{odake}. Therefore, apart from an
overall constant, $P_N(x)$ are the exceptional Laguerre polynomials $\hat{P}_{\ell,n}(x^2;g)$ with degree $2n+2\ell$ in $x$ \cite{odake}. This implies $N=2(n+\ell)$, which when substituted in \eqref{energy1} gives the energy eigenvalue expression to be
\begin{equation}
 E=4n.   \label{energy}
\end{equation}
With $N=2(n+\ell)$, it is easy to verify that  \eqref{fdiff}, coincides with the differential equation \cite{odake2} for the exceptional $X_{\ell}$ Laguerre polynomials
\begin{eqnarray}
 \partial_{xx}P_N(x)+2\left(-x+\frac{g+\ell}{x} -
\frac{\partial_x\xi_{\ell}(x^2;g)} {\xi_{\ell}(x^2;g)}\right)\partial_xP_N(x)  \nonumber\\  \left(4n-4\ell +4x\frac{\partial_x\xi_{\ell}(x^2,g+1)}{\xi_{\ell}(x^2,g)} \right)P_N(x)=0.
\end{eqnarray}
In addition, substituting the meromorphic form of the momentum function from \eqref{qmfmero} in \eqref{qmf}, we obtain the unnormalized wave function as
\begin{equation}
\Psi_n(x)= \frac{x^{(g+\ell)}\exp(-\frac{1}{2}x^2)}{\xi_{\ell}(x^2;g)}
\hat{P}_{\ell,n}(x^2;g),    \label{wf}
\end{equation}
which agrees with the known results \cite{odake}. Thus, we have obtained the energy eigenvalues and eigenfunctions for the new exactly solvable potentials, using the singularity structure of the momentum function. In addition to the fixed poles, $q_{\ell}$ has  $2n$ moving poles on the real line and $2\ell$ moving poles off the real line. The point at infinity turns out to be an isolated singularity.\\

\noindent
{\bf Singularity structure}\\

Here, we compare the singularity structure of the momentum function associated
with these models to those of the QES and the conventional ES models. We see
that the differences among these three models are due to the moving pole
structures. For all the models, the number of  moving poles located on the
real line is consistent with the oscillation theorem. For the conventional ES
models there were no moving poles off the real line. For the QES models,
moving poles off the real line were found. The number of these poles varied
from one QES level to the other, keeping the total number of moving poles
(located at both real and complex locations) fixed for all solvable
states. Compared to this, the new ES models have moving poles off the real
line, but their  number is fixed, equal to $2\ell$, for all the energy
levels. Thus, we see that the  singularity structure of the newly constructed
potentials is neither completely like the ES models nor the QES models. 

\noindent
\section{Exactness of SWKB quantization condition}
In an earlier paper \cite{bhalla}, the information on the singularity structure
of the QMF for several models, was used to account for the exactness of the SWKB quantization
condition,
\begin{equation}
\frac{1}{\pi}\int_{x_1}^{x_2} \sqrt{[E-W^2(x)]} dx= n, \label{swkb}
\end{equation}
where $x_1$ and $x_2$ $(x_1<x_2)$  are the turning points, which are the real
roots of $\sqrt{E-W^2(x)}$. For simplicity we denote the integrand $\sqrt{E-W^2(x)}$ as
$p_{SWKB}$ and put $\hbar =1$.  The integrals such as \eqref{swkb} can be turned into a contour integral (see for example \cite{kapoor}), and for \eqref{swkb} we get
\begin{equation}
I_C \equiv \frac{1}{2\pi }\oint_C  p_{{\scriptstyle SWKB}} dx= n, \label{oswkb}
\end{equation}
where $C$ is the contour enclosing the branch cut of $p_{SWKB}$ from $x_1$ to $x_2$. This integral can be
computed in terms of the contributions of the singularities outside the contour $C$. The residues of $p_{SWKB}$ are double valued and the right values are picked using the boundary condition, in the limit $E \rightarrow 0$,
$p_{SWKB} \rightarrow p$, since $W(x)= -\psi^{\prime}_0(x)/\psi_0(x)$. For the cases studied earlier in \cite{bhalla}, the residues and the locations of the singularities outside the contours $C$ of the two integrals, \eqref{oswkb} and
\eqref{exq} matched identically and hence application of complex integration techniques gave same answers. 


In the present case, we first list out the location of the poles of $p_{SWKB}$
outside the contour $C$ and corresponding residues in table 1 below.
\begin{table*}[h]
\begin{center}
\begin{tabular}[h]{|c|c|}
\hline
location & residue\\
\hline
$x=0$ & $ -i(g+\ell)$\\
\hline
$2\ell$ singularities  & $ -i $ \\
located at & \\
the zeros of & \\
 $\xi_{\ell}(x^2,g+1)$& \\
\hline
$2\ell$ singularities & $ i$ \\
located at & \\
the zeros of & \\
 $\xi_{\ell}(x^2,g)$&\\
\hline
$x=\infty$ &$ \frac{iE}{2} +i(g+\ell)$\\
\hline
\end{tabular}
\caption{Pole structure of $p_{SWKB}$ and $p(x)$.}
\end{center}
\end{table*}
 It can be easily seen that the poles and their residues are identical to those of the
QMF. However in this case, $p_{SWKB}$ has additional branch points as illustrated in figure 1, for the case of $l=1$ and $g=1$.   
\begin{figure}[h]
\centering 
{\includegraphics[scale=0.4]{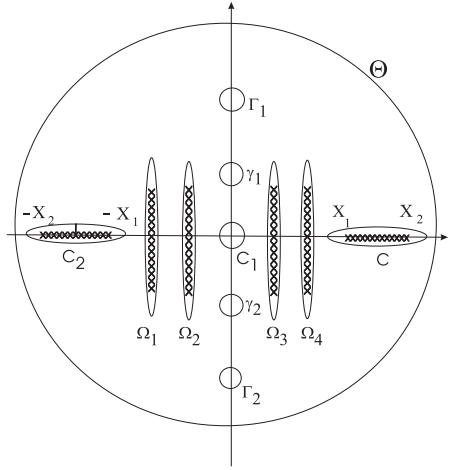}}
\caption{ The figure displays the singularities and branch cuts of
  $\sqrt{E-W^2(x)}$, for $g=1$ and $l=1$ and the contours enclosing them. All the contours here are oriented in the anti-clockwise direction.}
\end{figure}
In  figure 1, the contour $\Theta$ encloses all the poles and branch cuts of
$p_{SWKB}$. The contour $C_1$ encloses the  origin  and the contour $C_2$
encloses the branch cut between $-x_1$ and $-x_2$, which is equal to the
integral in \eqref{oswkb} by symmetry. The contours $\gamma_m,\, \Gamma_m $
enclose the simple poles 
corresponding to the $2\ell$ zeros of  $\xi_{\ell}(x^2,g)$ and
$\xi_{\ell}(x^2,g+1)$ respectively. Finally, the contours $\Omega_i$ enclose
the branch cuts of $p_{SWKB}$  located off the real line.  By writing the left hand side of \eqref{oswkb} in terms of contour integrals enclosing the poles and branch cuts of $p_{SWKB}$ located outside $C$ gives\\
\begin{eqnarray}
I_C &=& \frac{1}{2\pi}\oint_{\Theta} p_{SWKB}dx -\frac{1}{2\pi}\left(\oint_{C_1}  p_{SWKB}\, dx +  \oint_{C_2}
p_{SWKB}\, dx \right.\nonumber \\&& + \left.\sum_{m=1}^{2\ell} \left[\oint_{\gamma_{m}} p_{SWKB}\,dx +
	\oint_{\Gamma_{m}} p_{SWKB}  \, dx\right] + \sum_{i=1}^{4\ell}
\oint_{\Omega_i} p_{SWKB}\, dx \right).  \label{int} 
\end{eqnarray}
We show that the contribution of the branch cuts of $p_{SWKB}$ enclosed by
contours $\Omega_i$ should vanish, for the SWKB condition to be exact. For
this purpose we will evaluate the different integrals, except the ones around
the contours $\Omega_i$ in the left hand side of \eqref{int}.  The integral
around contour $\Theta$ is equal to $2\pi i(\text{ residue 
at x equal to } \infty)$, calculated using the rule \cite{kapoor} 
\newpage
\begin{eqnarray}
\text{Residue of } \sqrt{E-W^2(x)} && \text{ at  z equal to } \infty  = - \text{ Coefficient of t in the } \nonumber \\ &&\text{expansion  of }\sqrt{E-W^2(1/t)} \text{ for small t}.
\end{eqnarray}
By using $\frac{\partial_x\xi_{\ell}(x^2;g+1)} {\xi_{\ell}(x^2;g+1)}
=\sum_{i=1}^{2\ell}\frac{1}{x-a_i}$ and $\frac{\partial_x\xi_{\ell}(x^2;g)}
{\xi_{\ell}(x^2;g)}= \sum_{j=1}^{2\ell}\frac{1}{x-b_j}$, one obtains
 \begin{equation}
 \sqrt{E-W^2(1/t)}=\frac{1}{t}\sqrt{Et^2-\left(1-(g+\ell)t^2-t\sum_{i=1}^{2\ell}\frac{t}{1-ta_i}+t\sum_{j=1}^{2\ell}\frac{t}{1-tb_j}\right)^2},
 \end{equation}
 \begin{equation}
 \equiv \pm \frac{i}{t}\sqrt{1-Et^2-2(g+\ell)t^2-\sum_{i=1}^{2\ell}\frac{t^2}{1-ta_i}+\sum_{j=1}^{2\ell}
 \frac{t^2}{1-tb_j}+   H(t)},
\end{equation}
 where $H(t)$ denotes all the other higher order terms. Expanding the square
 root, one obtains
\begin{equation}
 \sqrt{E-W^2(1/t)} \equiv \pm \frac{i}{t}\left( 1-\frac{1}{2}(E+2(g+\ell)^2)t^2 +\ldots,  \right) 
\end{equation} 
which gives the residue at infinity to be $\pm i(\frac{E}{2} + (g+\ell))$. The
application of the boundary condition $E \rightarrow 0$, $p_{SWKB} \rightarrow
p$, allows us to choose the right value of the residue  to be $ i(\frac{E}{2}
+(g+\ell))$. Therefore
\begin{eqnarray}
 \frac{1}{2\pi}\oint_{\Theta} p_{SWKB}dx &=& -i\left(i\frac{E}{2}+i(g+\ell)\right) \nonumber\\
&=& \left(\frac{E}{2}+(g+\ell)\right)
\end{eqnarray}
 The other method to calculate the integral around $\Theta$ is to use
the mapping $x=1/t$, which  maps the point at infinity in the $x$ plane to the
origin in the $t$ plane. The coefficient of $1/t$ in the Laurent expansion of the
integrand in powers of $t$, will be the residue and is same as the value
calculated above.  
 
    The residues at the other singular points, listed in  table 1, can be obtained by using  simple complex variable techniques. Now with all the residues known, application of  the Cauchy residue theorem gives,
\begin{equation}
I_C=\frac{E}{2} + (g+\ell) - \left((g+\ell) + I_C + [-2\ell + 2\ell] +\frac{1}{2\pi}\sum_{i=1}^{4\ell}\oint_{\Omega_i} p_{SWKB} dx \right). \label{temp2}
\end{equation}
Simplifying the right hand side gives
\begin{equation}
2 I_C=\frac{E}{2} - \frac{1}{2\pi}\sum_{i=1}^{4\ell}\oint_{\Omega_i} p_{SWKB} dx.  \label{emp3} 
\end{equation}
If $E$ coincides with the $n^{th}$ eigenvalues given in \eqref{energy}, we obtain
\begin{equation}
 I_C= n - \frac{1}{4\pi}\sum_{i=1}^{4\ell}\oint_{\Omega_i} p_{SWKB} dx.  \label{emp10} 
\end{equation}
Therefore, it is obvious that if the sum of the contribution of the branch cuts off the real line vanishs, we obtain
\begin{equation}
 \frac{1}{2\pi}\oint_C p_{SWKB}=n
\end{equation}
Thus vanishing of the contribution of the branch cuts off the real line turns out to be a necessary
and sufficient condition for the exactness of the SWKB rule. 

  We point out here that the calculation of the integral appearing in
  \eqref{exq}, in terms of $p$, also proceeds in a similar fashion. The only
  difference being that the QMF does not have branch points or moving branch
  point singularities in the complex plane. Therefore this integral receives
  contribution only from the fixed poles corresponding to the  potential, the $n$
  moving poles located symmetrically between $-x_1$ and $-x_2$ and from the point
  at infinity. As the table suggests the contribution from these poles, for both
  integrals is identical. Therfore it is obvious that the SWKB condition is
  exact if the contribution from the branch cuts off the real line vanishes,
  when the energy is an eigenvalue. 

In \cite{khbook},\cite{khp}  it has been proved that the SWKB condition is exact for all the  conventional ES, translationally SIPs. The  proof makes use of the property of shape invariance,
where the assumption that  $W^2(x)$ is of
$O(\hbar^{0})$  while $\hbar\partial_xW(x)$ is of $O(\hbar)$ was the key point. However, unlike the conventional SIPs with translation, for the new SIPs with translation, $W(x)$ is in fact a complicated function
of $\hbar$. Thus the derivation in \cite{khbook}, \cite{khp} about the exactness of SWKB for SIPs with
translation is strictly not applicable for the newly discovered SIPs with translation. Our present study of the exactness of the SWKB rule suggests that the disappearance of the contribution of the branch cuts off the real line, when $E$ is equal to the one of the eigenvalue is possibly linked to the property of shape invariance.

\noindent
\section{Conclusions}
In this study, we have analyzed the new ES shape invariant potentials obtained
by deforming the radial oscillator using the QHJ formalism and obtained the
expressions for the eigenvalues and eigenfunctions. We have investigated the
singularity structure of the QMF and compared it with those of the conventional ES
and QES models studied earlier. The QMF for these potentials has $2n$ real and $2\ell$ complex moving poles, with $E=4n$ giving the energy eigenvalue and $\ell$ fixing the potential. We found that the singularity structure of these models is neither completely like the ES models nor the QES models. The exactness of the SWKB rule has been shown to be equivalent to vanishing of the contribution of the branch cuts off the real line.  Lastly, we point out that the above results can be extended to the
other families of infinite number of potentials obtained by deforming the
Darboux-P\"oschl-Teller potentials \cite{odake}. \\

\noindent
{\bf Acknowledgments} S S R thanks the Centre for Advanced Studies (CAS), School
of Physics, University of Hyderabad, Hyderabad and the Department of Science and
Technology (DST), India (fast track scheme (D. O. No: SR/FTP/PS-13/2009)) for financial support.\\


\noindent
\section{ References}

\begin{thebibliography}{00}


\bibitem{odake} Odake S and Sasaki R 2009 {\it Phys. Lett. B} {\bf679} 414 (arXiv:0906.0142).

\bibitem{odake2} Sasaki R, Tsujimoto S and Zhedavov A 2010 {\it J. Phys. A: Math. Theor} {\bf 43} 315204 (arXiv:1004.4711).

\bibitem{dar} Darboux G, Acad C R  1882 {\it Paris} {\bf 94} 1456.

\bibitem{crum} Crum M M 1955 {\it Quart. J. Math. Oxford Ser.(2)} {\bf 6} 121 (arXiv:9908019).

\bibitem{nick1} G\'omez-Ullate D, Kamran N and Milson R 2010 {\it J Approx Theory} {\bf 162} 987 (arXiv:0805.3376).

\bibitem{nick2} G\'omez-Ullate D, Kamran N and Milson R 2009 {\it J. Math. Anal. Appl.} {\bf 359} 352 (arXiv:0807.3939).

\bibitem{lea1} Leacock R A and Padgett M J 1983 {\it Phys. Rev. D} {\bf 28} 2491.

\bibitem{lea2} Leacock R A and Padgett M J 1983 {\it Phys. Rev. Lett.} {\bf50} 3.

\bibitem{bhallaAJP} Bhalla R S, Kapoor A K and Panigrahi P K 1997 {\it Am. J. Phys.} {\bf 65} 1187.

\bibitem{es} Sree Ranjani S, Geojo K G, Kapoor A K and Panigrahi P K 2004 {\it Mod. Phys. Lett. A} {\bf 19} 1457.

\bibitem{qes} Geojo K G , Sree Ranjani S and Kapoor A K 2003 {\it J. Phys. A: Math. Gen.} {\bf36} 4591.

\bibitem{the} Sree Ranjani S 2005 {\it Quantum Hamilton - Jacobi solution
  for spectra of several one dimensional potentials with special
  properties} Thesis submitted to the University of Hyderabad (arXiv:0408036).

\bibitem{khareAJP} Dutt R, Khare A and Sukhatme U P 1988 {\it Am. J. Phys.} {\bf 56} 163.

\bibitem{seetha} Raghunathan K, Seetharaman M and Vasan S S 1987 {\it Phys. Lett. B} {\bf 188} 351.

\bibitem{bhalla} Bhalla R S, Kapoor A K and Panigrahi P K 1996 {\it Phys. Rev. A} {\bf 54} 951.


\bibitem{khbook} Cooper F, Khare A and Sukhatme U P 2001 {\it Supersymmetric quantum mechanics}( Singapore: World Scientific Publishing Co. Ltd.)

\bibitem{quesne} Quesne C 2008 {\it J. Phys. A: Math. Theor.} {\bf41} 392001.

\bibitem{asim} Bougie J, Gangopadhyaya A, and Mallow J V 2010 {\it Phys. Rev. Lett.} {\bf 105} 210402.


\bibitem{morse} Morse P M, Feshbach H 1953 {\it Methods of theoretical physics (part 1)} (New York: McGraw-Hill Book Company Inc.) pp. 784-785.

\bibitem{kapoor} Kapoor A K  2011 {\it Complex variables, principles and problem sessions} (Singapore: World Scientific Publishing Co. Ltd.) chap. 6, sec. 8.

\bibitem{khp} Dutt R, Khare A, Sukhatme U P 1986 {\it Phys. Lett. B} {\bf 181} 295.

\end{thebibliography}
\end{document}